\documentclass[aps,floatfix,superscriptaddress,notitlepage,showkeys]{
revtex4}
\usepackage{amsmath,amssymb,amsfonts,graphics,graphicx,dcolumn,bm,enumerate}
\usepackage{comment,natbib,appendix}
\usepackage[ddmmyyyy,hhmmss]{datetime}
\usepackage{multirow,color}
\usepackage{adjustbox}
\usepackage{float}

\usepackage{amsthm}
\usepackage{epsfig}
\usepackage{natbib}
\usepackage{hyperref}
\usepackage{graphicx}
\usepackage{epstopdf}

 \newcommand{\tcs}
{\affiliation{TCS Research, New Delhi, India}}
 \newcommand{\jnu}
{\affiliation{School of Computational and Integrative Sciences, Jawaharlal Nehru 
University, New Delhi-110067, India}}

\begin{document}

\title{A complex network analysis of ethnic conflicts and human rights violations}

\author{Kiran Sharma} 
\email[Email: ]{kiran34\_sit@jnu.ac.in} 
\jnu
\author{Gunjan Sehgal}
\email[Email: ]{sehgal.gunjan@tcs.com} 
 \tcs
\author{Bindu Gupta}
\email[Email: ]{bindu.gupta2@tcs.com} 
 \tcs
\author{Geetika Sharma}
\email[Email: ]{geetika.s@tcs.com}
 \tcs
\author{Arnab Chatterjee}
\email[Email: ]{arnab.chatterjee4@tcs.com} 
 \tcs
\author{Anirban Chakraborti}
\email[Email: ]{anirban@jnu.ac.in}
 \jnu
\author{Gautam Shroff}
\email[Email: ]{gautam.shroff@tcs.com} 
 \tcs


\begin{abstract}
News reports in media contain records of a wide range of socio-economic and 
political events in time. Using a publicly available, large digital database of 
news records, and aggregating them over time, we study the network of ethnic conflicts 
and human rights violations. Complex network analyses of the events and the involved 
actors provide important insights on the engaging actors, groups, establishments and 
sometimes nations, pointing at their long range effect over space and time. We find 
power law decays in distributions of actor mentions, co-actor mentions and degrees 
and dominance of influential actors and groups. Most influential actors or groups 
form a giant connected component which grows in time, and is expected to encompass 
all actors globally in the long run.  We demonstrate how targeted removal of actors 
may help stop spreading unruly events. We study the cause-effect relation between 
types of events, and our quantitative analysis confirm that ethnic conflicts lead to 
human rights violations, while it does not support the converse.
\end{abstract}

\maketitle

% \noindent Please note: Abbreviations should be introduced at the first mention in the main text – no abbreviations lists. Suggested structure of main text (not enforced) is provided below.

\section{Introduction}
The quantitative analyses of the gigantic amount of data associated with human social 
conditions have gained much momenta in the recent years, especially through the 
multidisciplinary tools and approaches. The traditional theories of social science 
together with complex network 
analysis~\cite{newman2006structure,Albert:2002,barabasi2016network} 
and addition of new tools and paradigms from several science disciplines like 
theoretical physics, applied mathematics, computer science, as well as psychology, 
have developed into what is presently known as computational social 
science~\cite{lazer09} and have helped uncover new patterns of social behavior, 
including social dynamics~\cite{Castellano:2009,Sen:2013}. Data made available 
digitally, has drawn attention of researchers across disciplines who have 
collaborated and contributed in their own ways to scientifically analyze 
and understand complex social phenomena in the recent years, previously not known to 
the present scale of detail.

Social behavior, conditions, events, organizations, which have given shape to the 
history of human civilization, have often been positive or constructive -- building 
societies, shaping cultures, and at other times negative or destructive -- conflicts, 
wars and battles followed by reorganization of countries or nations. Human 
civilization has evolved in a complicated manner and created a variety in ethnic 
characters and cultures, in languages, beliefs and customs.
It is widely understood~\cite{toft2005geography} that historically, certain groups 
pay priority to their linguistic, cultural, racial, and religious ties of individuals 
within the groups, which are then passed down through generations. Ethnic violence 
arises when groups attempt to maintain their boundaries against pressure from 
historical enemies. 
In another possible scenario, when the authority of a multi-ethnic state declines, 
the central regime ceases to protect the interests of ethnic groups, creating a void 
in which ethnic groups start competing to establish and control a new regime that 
will protect their interests. Socio-political tension in such a situation is likely 
to incite violence.

Ethnic heterogeneity is a characteristic feature observed in most countries and 
regions worldwide. With time, these variations often reduce locally, where 
socio-economic and political forces are less dominant in favor of ethnic mixing. 
However, the course of political history along with that of religion and culture 
often encounters conflicts at various scales. Ethnic conflicts are comparatively 
frequent in certain regions and rare elsewhere. Ethnic heterogeneity does not 
necessarily breed war and its absence does not ensure peace. Even today, ethnic wars 
continue to be globally the most common form of armed conflicts, but the mechanisms 
that lead a society down the path of ethnic conflict are yet to be fully understood. 
What is intriguing, is the connection between democratization and the occurrence of 
ethnic conflict. While stable democracies are
unlikely to wage war with other democracies, a country that is 
socio-politically unstable may very well find conflicts between groups with opposing 
interests~\cite{vorrath2007linking}. Human rights are internationally agreed values, 
standards or rules regulating the conduct of states towards their own citizens and 
towards non-citizens. Interestingly, the violation of human rights appears to be 
more 
associated with ethnic conflicts than abuses of economic and social 
rights~\cite{thoms2007human}.
Other political, economic and social preconditions may also influence the causes of 
ethnic conflicts, and the conscious promotion by the political actors of any 
polarizing dimension based on these factors is sufficient to lead to conflict. It is 
often seen that changes in 
political regime can lead to conflicts which are often ethnic in various parts of 
the 
world, sparking human rights violations. Therefore, spatio-temporal analyses of 
regional conflict formations and political dynamics, and the statistical studies of 
the different variables are important.

Temporal data of events like human to human communications and physical 
contacts/proximity has been studied in recent years (see e.g., Holme and Saram\"{a}ki
\cite{holme2013temporal}) in connection to study of epidemics and 
contagion, spreading of information using mobile phone communication 
data~\cite{onnela2007structure} as well as data from 
proximity sensors carried by human entities~\cite{cattuto2010dynamics}.

Here we provide a quantitative analysis of the scale and topology of 
ethnic tensions, related conflicts and violations of human rights, using the reports 
in digital media. We look at data from a 
publicly accessible database, which keeps  account of events from news available in 
media. We particularly focus on ethnic conflicts (EC) and human rights violations 
(HR) 
by suitably filtering keywords present in the digital 
text transcript of the news. With the availability of high precision data containing 
precise spatio-temporal information, one can look towards finding correlations 
between events, involved actors (individuals, groups, organizations or states), the 
geographical pattern of spreading of conflicts etc.
Although there have been studies on conflicts, wars and terror 
attacks~\cite{richardson1960statistics,clauset2007frequency,becerra2012natural,
chatterjee2017fat}, and speculation of ethnic conflicts using census data for 
segregated population~\cite{lim2007global}, 
a comprehensive study of the actors involved in ethnic conflicts 
and human rights violation has been lacking.
In this work, for the first time, we provide a quantitative and qualitative 
understanding of the relative activity of the actors, frequently engaging actor 
pairs, the network of actors, and give insights into the static and dynamical 
aspects of the actor network in a scenario involving ethnic conflicts and human 
rights violations. 

\section{Results}
GDELT Event Database~\cite{GDELT} contains the database of news articles from all 
over 
the world in several languages. Using a query, it is possible to extract data for 
events, about ethnic conflicts (EC) and human rights violations (HR) happening around 
the world, spanning over a large time scale. Each event data contains information 
such as the pair of \textit{actors} involved, a unique time stamp, names of 
individuals, organizations or groups, the location information of the event, as well 
as latitude, longitude data of actors and the event. The news event and subsequently 
the data serves as a proxy for the actual event and its intensity (in terms of number 
of reports). Although we do not analyze the actual news reports, we consider a report 
on EC to be an actual instance of ethnic conflict or its possibility, and similarly 
for HR. We analyzed $28,055$ events of EC and $36,470$ of HR for a 15 year period 
(2001-2015).
%%%%%%%%%%%%%%%%%%%%%%%%%%%%%%%%%%%%
\begin{figure}[h]
\centering
\includegraphics[width=16.0cm]{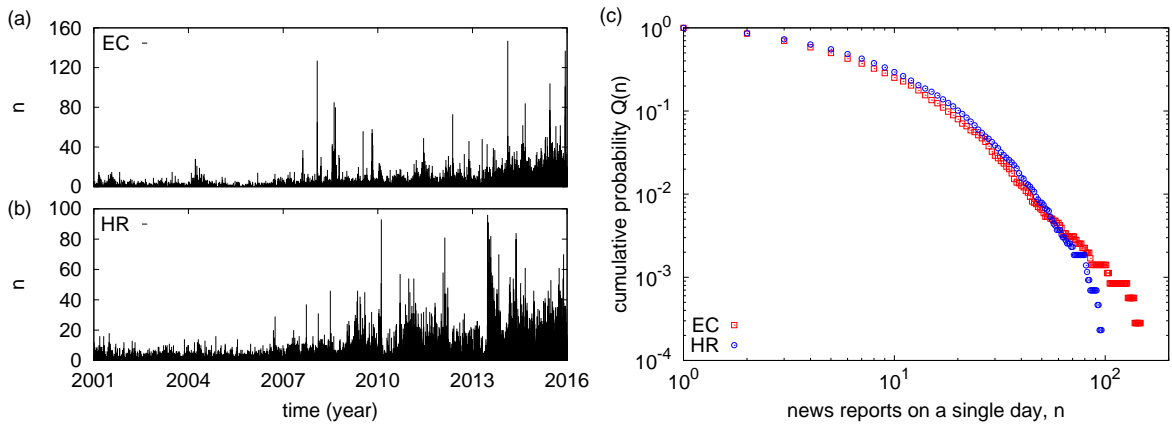} %\vskip 1.0cm
\caption{The time sequence of the number of events $n$ reported daily 
for (a) EC, (b) HR  during 2001-2015.
(c) The cumulative probability (CCDF) $Q(n)$ that $n$ or more events are reported 
on a particular day.
The data for EC seems to fit well to a power law for the largest values, with decay 
exponent around $2.54 \pm 0.03$, HR fits well to a stretched exponential 
($\exp[-an^b]$ with 
$a=0.44 \pm 0.12$ and $b=0.62 \pm 0.01$). }
\label{fig:burst}
\end{figure}
%%%%%%%%%%%%%%%%%%%%%%%%%%%%%%%%%%%%%%%%

The number of news entries  $n$ per day is a stochastic variable, and often there is 
a burst of activity noted (Fig.~\ref{fig:burst}a,b). The amount of data cataloged 
is also observed to have 
increased since the late 2000's. The number of news entries/ reports on a single day 
$n$ has a broad distribution due to the large inter-day fluctuations in the number 
of reports and bursty nature of the data. 
Fig~\ref{fig:burst}c shows 
the complementary cumulative distribution function (CCDF) 
that a day has more than $n$ news reports, 
$Q(n)$. For EC, the 
distribution has a short lognormal body with a rather prominent power law tail 
(decay 
exponent close to $2.54 \pm 0.03$); for HR, the bulk of the distribution fits to a 
stretched 
exponential ($\exp[-an^b]$ with $a=0.44 \pm 0.12$ and $b=0.62 \pm 0.01$)
The details of the database, data extraction and different attributes are described 
in Supplementary Information. In our study, we focus on a relatively recent 
time 
span (2001-2015) and extract the data for actor pairs and geographical location of 
events from the database.

\subsection{Static properties of the network}
\begin{figure}[h]
\centering
\includegraphics[height=9.2cm]{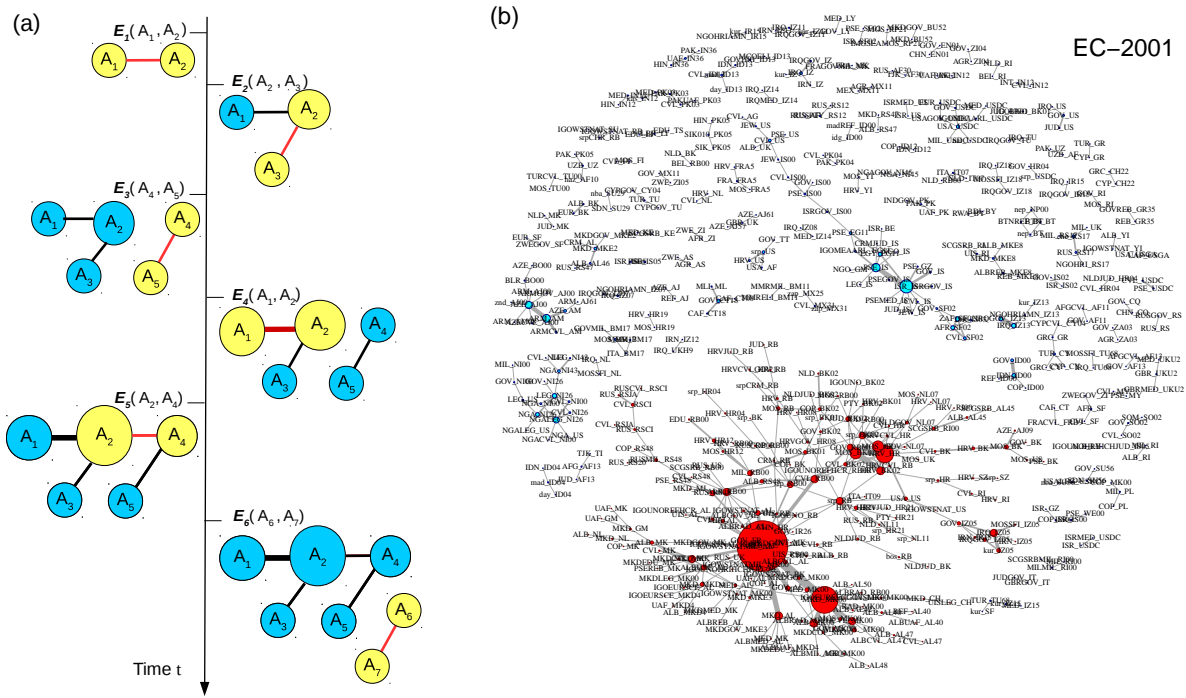} 
\caption{(a) The schematic visualization of events on a time-line and the subsequent 
evolution of the aggregated network of actors: Events appear sequentially. Event 
$E_1$ involving actors $A_1,A_2$ is followed by event $E_2$ involving $A_2,A_3$ and 
so on. The links depict the actors involved together in an event. 
New events (actors in yellow and links in red) add to actor mention (size of node) 
and actor co-mention (link weight; thickness of links).
(b) The network of actors for EC for 2001. Each actor is a 
node and any two actors ever involved in an event are connected by a link. The size 
of the nodes are proportional to the total number of mentions in the period, and the 
relative thickness of the links are proportional to the total number of co-mentions 
of the actors (link weights).
The largest connected component (giant component) is depicted in red.}
\label{fig:netviz}
\end{figure}

We construct the aggregate network over a given period of time, for the entire
15 year period (2001-2015) as well as for each of the 15 individual years, for the 
two datasets.
Each event is visualized as a link between the actors involved, which are the nodes, 
thus creating 
a network of actors connected by events.
The details of the construction is given in the Methods section. A schematic 
representation 
of the network construction
is given in Fig.~\ref{fig:netviz}a and a typical network is shown 
in Fig.~\ref{fig:netviz}b for EC data aggregated over the year 2001.

%%%%%%%%%%%%%%%%%%%%%%%%%%%%%%%%%%%%
%%%%%%%%%%%%%%%%%%%%%%%%%%%%%%%%%%%%%%%%
In our study, the temporal granularity of the data is one day. 
We extracted (i) the number of mentions $m$ of each individual actor and (ii) the 
number 
of co-mentions $w$ of an unique pair of actors (actor pair mentions),
as well as the number of unique actors $k$, one actor is involved with.
In terms of the network, $m$ measures node strength, $w$ the link weight and $k$ the 
degree of a node.
While $m$ and $k$  measure the importance, activity or visibility of a 
single actor, $w$ measures the involvement of an actor pair 
in inciting an event.
These quantities are measured for each year for 
the 15 year period (2001-2015) as well as through the whole period (aggregate over 
15 years).

\begin{figure}[h]
\centering
\includegraphics[width=17.0cm]{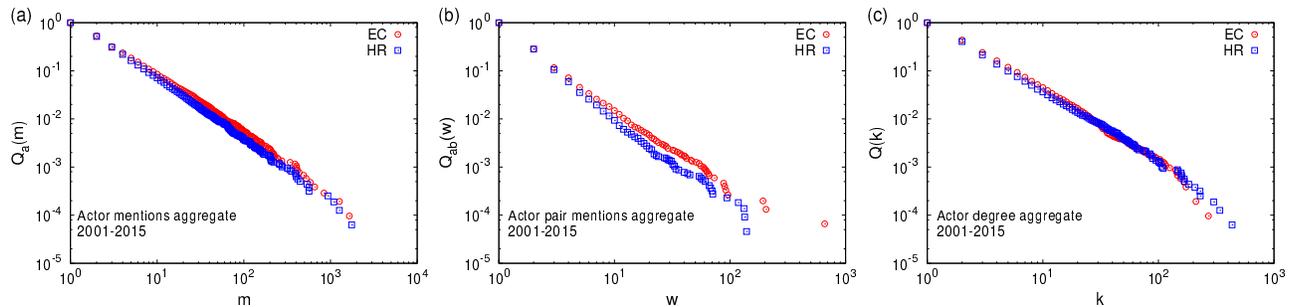}
\caption{Structural properties of the aggregated network in the period 2001-2015:
(a) Plot of the cumulative probability (CCDF) $Q_a(m)$ that an actor is mentioned 
at least $m$ times. Except for an exponential decay at 
the very end of the tail, most of the distribution has a the power law decay $Q_a(m) 
\sim m^{-\nu_1}$ with exponents $\nu_1$ as $1.23 \pm 0.01$ for EC and $1.28 \pm 
0.01$ 
for HR.
(b) Plot of the cumulative probability $Q_{ab}(w)$ that an actor pair is mentioned 
at least $w$ times. The tail of the distributions fit 
well to power laws $Q_{ab}(w) \sim w^{-\nu_2}$ with exponents as $1.58 \pm 0.01$ for 
EC and  $1.74 \pm 0.03$ for HR. 
(c) Plot of the cumulative probability $Q(k)$ that an actor is connected to $k$ 
others or more. The tail of the distributions fit well to 
power laws $Q(k) \sim k^{-\nu_3}$ with exponents $\nu_3$ as $1.52 \pm 0.01 $ for EC 
and 
$1.48 \pm 0.01$ for HR.}
\label{fig:allyr_cumcount}
\end{figure}

It is often perceived that certain actors frequently engage in reported events than 
others and the same is true for pairs of actors engaging in ethnic conflicts or 
human rights violations. 
We computed the complementary cumulative probability distribution (CCDF) for 
different quantities for the aggregated network in $2001-2015$ (refer to 
Fig.~\ref{fig:allyr_cumcount}a,b,c).            
%%%%%%%%%%%%%%%%%%%%%%%%%%%
%%%%%%%%%%%%%%%%%%%%%%%%%%%
For the actor mentions, the  probability that an actor is mentioned $m$ times or 
more  fits to a power law for the largest values, $Q_a(m) 
\sim m^{-\nu_1}$, with decay exponents $1.23 
\pm 0.01$ for EC and  $1.28 \pm 0.01$ for HR. The exponents are computed, along with 
the values of the standard errors (uncertainties) and significances, using Maximum 
Likelihood Estimates~\cite{Clauset:2009}.
The probability that an actor pair is mentioned $w$ times or more  
fits well to power laws $Q_{ab}(w) \sim w^{-\nu_2}$ with exponents 
$\nu_2$ as $1.58 \pm 0.01$ for EC and 
$1.74 \pm 0.03$  for HR data.
The \textit{actor mentions} and \textit{actor pair mentions} are 
respectively the \textit{node weights} and \textit{edge weights} in network 
terminology.
The probability that an actor was involved with $k$ or more actors during the time 
span  (degree of the actor)
fit well to power laws $Q(k) \sim k^{-\nu_3}$ with exponents $\nu_3$ as 
$1.52 \pm 0.01$ for EC and  
$1.48 \pm 0.01$ for HR data.
For the individual years, due to less aggregation, the counts are less and 
data naturally seems noisy. However, the data still seem to exhibit the power law 
tail in the probability distributions (see Supplementary Fig.~S7). The 
fitting exponents for the individual 
years are given in Supplementary Table~S1.
The correlation between actor degree and mentions, as well as the relationship 
between the power law exponents $\nu_1$ and $\nu_3$ are shown in Supplementary 
Fig.~S8.

The above results quantitatively characterize the heterogeneity in the activity of 
actors, while most actors are relatively less active: the power law distributions 
for 
actor mentions $Q_a$ indicate that there are a significant few who constantly engage 
in ethnic conflicts and human rights violations issues. The power law distributions 
in actor pair mentions $Q_{ab}$ indicate similar characteristic for pairs of actors. 
The broad degree distributions $Q(k)$ are indicative of the fact that the number of 
actors engaging with very large number of actors are also significant.

%>>>>>>>>>>>>>>>>>>>>>>>>>>>>>>>>>>>>>>>>>>>>>>>>>>>>>>>>>
\subsection{Clusters}
The aggregated network is found to be composed of several disconnected components or 
`clusters'. Physically this means that the actors constituting one cluster have 
never 
been involved with any actor from a different cluster. For both sets, the 
largest connected component is $10^2-10^3$ times the smaller clusters, which are 
large in number (see Supplementary Table.~S2). 
In fact, the largest cluster $s_1$ grows superlinearly with the size of 
the network $N$, as we found (from the data in Supplementary Table.~S2), $s_1 \sim 
N^\delta$, with $\delta=1.17\pm 0.01$ for EC and $1.19 \pm 0.02$ for HR 
(Supplementary 
Fig.~S9).
This tells us that the 
fraction of nodes in the largest cluster grows with the total size of the network 
quite fast so  
that eventually the fraction of nodes outside the largest cluster will be 
negligible.
We computed the CCDF  $Q(s)$ of 
cluster size $s$ and find that it roughly has a power law decay for components 
except 
the largest one, with a decay exponent roughly close to 3 
(see Fig.~\ref{fig:clus_size1}). The CCDF for individual years has been shown in 
Supplementary Fig.~S10.
The largest clusters in both sets have a slowly decaying clustering coefficient 
$\langle C(k) \rangle$ with degree $k$ (see Supplementary Fig.~S11).
%%%%%%%%%%%%%%%%%%%%%%%%%%%
\begin{figure}[h]
\centering
\includegraphics[width=8.0cm]{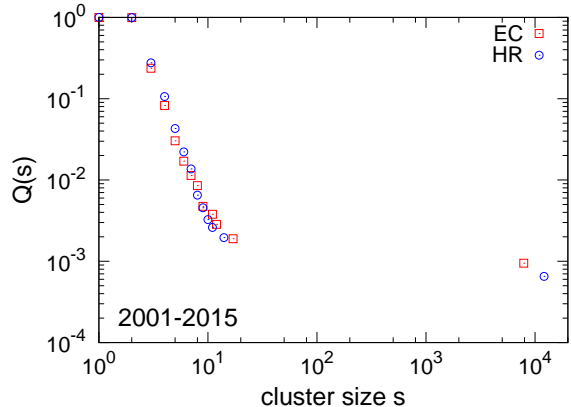} %\vskip 1.0cm
\caption{Plot of the cumulative probability (CCDF) $Q(s)$ that there is a cluster of 
size 
larger than $s$, for data aggregated over the period 2001-2015.
The size of the largest cluster is very large compared to the rest.
}
\label{fig:clus_size1}
\end{figure}
%%%%%%%%%%%%%%%%%%%%%%%%%%%

\subsection{Network growth properties}
In order to explore the dynamics of the process that leads to the broad 
distribution 
of the mentions and degrees, we investigate the dynamics of growth for these 
quantities over the span of 15 years (2001-2015). For good statistics, we ranked the 
actors according to the number of mentions and degrees, and analyzed the data for 
the top 10 actors.

We first measured the growth of the degrees and mentions for the top 10 actors of EC 
and HR data, with respect to the time $t^*$ when they were first mentioned. The 
long time behavior (asymptotic) indicates a growth law of $(t-t^*)^\beta$ with 
$\beta \simeq 3$.
%%%%%%%%%%%%%%%%%%%%%%%%%%%%%%%%%%%%
\begin{figure}[h]
\centering
\includegraphics[width=17.0cm]{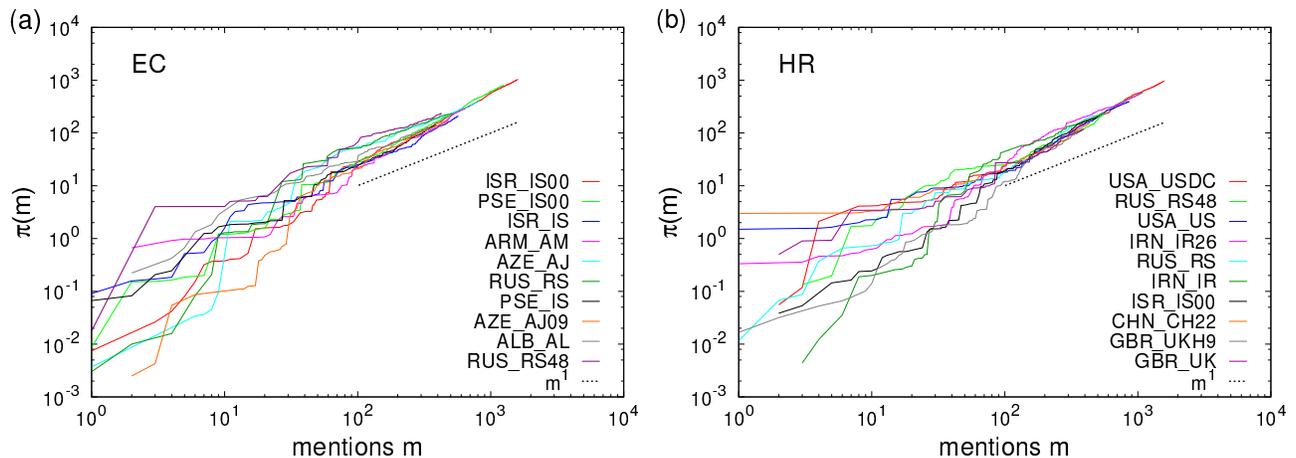}
\caption{Cumulative growth rates $\pi(m)$ for mentions $m$ for (a) EC and (b) HR 
datasets. 
The curves asymptotically fit to  $\pi(m) \sim  m^a$ with $a > 1$. The precise 
fitting exponents 
are given in Supplementary Table.~S4.}
\label{fig:pik_men}
\end{figure}
%%%%%%%%%%%%%%%%%%%%%%%%%%%%%%%%%%%%%%%

To extract the rate of growth for a quantity $x$, we computed $\Pi(x) = \frac{\Delta 
x}{\Delta t}$. $\Pi(x)$ in real data turns out to be very noisy, so we computed the 
cumulative integral $\pi(x) = \int_0^x \Pi(x^\prime) d x^\prime$ which turns out to 
be less noisy. We observe that data for degree $\pi(k)$ has a shorter span and is 
noisier than that for mentions $\pi(m)$. 
The plots suggest that both $\pi(m)$ and $\pi(k)$ are superlinear 
functions of their respective arguments, i.e., $\pi(x) \sim  x^\alpha$ with $\alpha 
\sim 
1.2-1.4$ (see Fig.~\ref{fig:pik_men} 
for mentions and Supplementary Fig.~S12 for degree). 
Thus the asymptotic growth rate $\Pi(x) \sim x^{\alpha-1}$ is still  weakly 
dependent on the respective arguments. We thus identify that the growth rate of the 
system (network) is not independent of the size of the node (degree or mentions).
The growth exponents are computed using Maximum Likelihood 
Estimates~\cite{Clauset:2009} and tabulated in Supplementary 
Table.~S4.  
A superlinear growth rate implies that the network will, in the long run,
consist of a very large `condensate' with a small fraction of nodes in isolated 
smaller clusters,
which is consistent with our analysis of the cluster size distribution.

\subsection{Tolerance to attack and failure}
\begin{figure}[h]
\centering
\includegraphics[width=17.0cm]{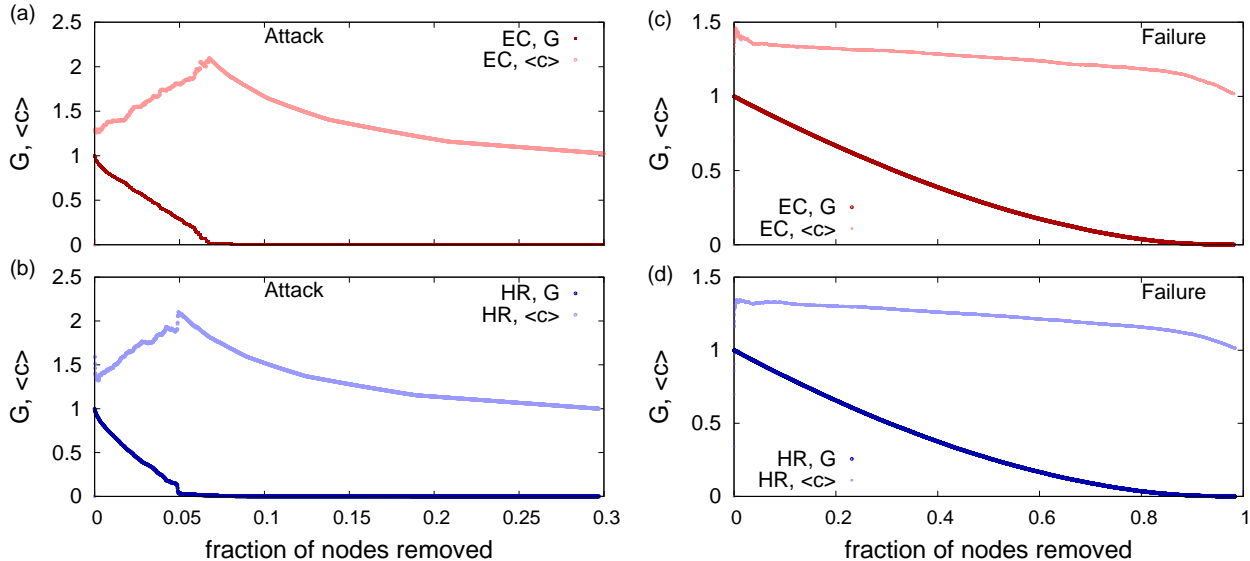} %\vskip 1.0cm
\caption{The structure of the network under attack: 
Nodes are removed in the sequence of
their degrees starting from the highest degree. The plot shows the behavior of the 
giant component $G$
(fraction of nodes in the largest connected component) and the average number of 
nodes in the clusters other
than the giant component $\langle c \rangle$, with increasing fraction of removed 
nodes, for (a) EC and (b) HR.
The structure of the network under random failure: Nodes are removed randomly.
Results are shown for (c) EC and (d) HR networks.
The networks are destroyed very quickly by targeted node removal (attack), compared 
to random node removal (failure).
The results are for networks aggregated over 2001-2015.}
\label{fig:attack}
\end{figure}

We also study how the network breaks down under attack, in order to investigate
the possibility of preventing unruly events to spread~\cite{albert2000error}.
The largest connected component of the network is subjected to targeted attack by 
removal of the most connected nodes.
We start by removing the node with the highest degree, followed by the next highest 
and so on. This results in rapid fragmentation or destruction of the network. 
We compute the fraction of nodes $G$ present in the largest cluster, which is 
observed to decrease very quickly (Fig.~\ref{fig:attack}a). In fact, the network of 
actors can be destroyed by targeted attack just by removing much less than 10\% of 
the nodes compared to when randomly selected nodes are removed (random failure) one 
after another (Fig.~\ref{fig:attack}b).
This exercise indicates that targeted intervention may help stop spreading of ethnic 
conflicts and human rights violations.
%%%%%%%%%%%%%%%%%%%%%%%%%%%%%%%%%%%%
%%%%%%%%%%%%%%%%%%%%%%%%%%%%%%%%%%%%%%%

\subsection{Measuring causality}
Considering that we have time series data for the number of mentions of ethnic 
conflicts and human rights violations from news articles it would be interesting to 
study if there is a clear cause and effect relation between these events with time. 
To determine the causal effect purely from the observations of the past data, we 
apply Granger Causality~\cite{granger1969investigating} which estimates the causal 
relationship by observing the changes in the distribution of the variables over 
time. 

Let us consider two random variables depicting counts of EC and HR. 
To say that EC causes HR, Granger causality computes a regression of variable HR on 
the past values of itself 
and the past values of EC and then tests the significance of coefficient estimates 
associated with EC.
We consider a bivariate linear autoregressive model on EC and HR, and assume the 
L.H.S. to be dependent on the history of EC and HR,
\begin{equation*}
HR_t  =  a_0 + a_1 EC_{t-1}+ \ldots + a_h EC_{t-h} + b_1 HR_{t-1}+ \ldots +b_h 
HR_{t-h} + E_t
\end{equation*}
where $h$ is the maximum number of lagged observations (for both EC and HR). The 
coefficients $a_i$, $b_i$ are the contributions of each lagged observation to the 
predicted value of $EC_{t-i}$ and $HR_{t-i}$ respectively and $E_t$ is the 
prediction error.

We set up a null hypothesis and to test the significance of the coefficients, 
compute the $p$ value.  If the $p$-value is less than $0.05$, one can reject the 
null hypothesis.
We tested the counts for year wise as well as month wise mentions. 
Comparing the $p$ values for both cases, we found that: (a) one cannot conclude that 
HR causes EC, but (b) EC causes HR. 
Hence, we can confirm that ethnic conflicts cause human rights violations, while
human rights violations are not responsible for ethnic conflicts.
The details of the analysis are provided in the Supplementary Information.

\section{Discussions}
News reports serve as a reasonable proxy for the importance and intensity of events. 
The intensity is reflected by the number of reports and lingering span of time 
through which the reports follow.
Our study focuses on events that pointed at ethnic conflicts and human rights 
violations.
The GDELT data is unique in the sense that it records events and actors involved in 
them. 
The U.S. government had recently funded in a large-scale project, the Integrated 
Conflict Early Warning System (ICEWS)~\cite{Ward2013},
which makes use of quantitative data and statistical methods in order to forecast events of
political instability, which include international and domestic crises, ethnic and 
religious violence, rebellion and insurgency. We have chosen the GDELT data instead 
of the ICEWS data to get a more global perspective~\cite{Arva2013,leetaru2013gdelt}.
The frequencies of mentions of actors and their co-mentions with others can 
be treated well as proxies for their importance and influence, as well as 
involvement with others.
The aggregate data enables us to construct a network of actors, and even finding 
disconnected groups.
In fact, our study reveals that most events are disconnected in very small 
clusters while very large clusters of frequently engaging actors exist.
One can study the geographical localization of events and clusters to procure 
detailed information regarding the context, intensity and growth pattern of 
events~\cite{futurework}.
Identifying important groups of actors, in terms of their intensities of 
activities, is important for possible intervention that may 
prevent the spread of such events. 
The probability distributions of actor mentions, co-actor mentions and the degree of 
an actor have power law tails for the largest values, indicating a strong 
self-organizing principle behind the events. 
The growth properties of individual actor nodes indicate that in the long run, very 
small fraction of disconnected clusters are left, while most of the actors belong to 
a giant connected component. The data on ethnic conflicts and human rights 
violations are found to be strikingly similar in terms of static and dynamic 
properties of the network, and even in terms of network stability against failure 
as well as targeted attack. This may point at a very high degree of correlation 
between events. In fact, using a causality analysis, we could quantitatively conclude 
that ethnic conflicts lead to human rights violations, while the reverse may not be 
true. 
These networks of ethnic conflicts and human rights violations reflect 
the negative aspects of human behavior and cooperation, and are certainly different 
from other social networks (friendship, collaboration, etc.) in the sense that there 
are not many triangles or closed communities, as  observed in the distribution of the 
average clustering coefficient with degree of the actors.
This detailed scientific study of the network structure, dynamics, function 
and resilience may help policy makers, specifically in cases where there is a need 
for preventing the spread of ethnic tensions and conflicts.
There are possibilities of similar analyses using data from online social media, 
e.g., 
\textit{Twitter}, etc. 

We have deliberately avoided mentioning the real names and other details because the 
main aim of the paper was to study the network properties and statistical 
regularities. Since we have just considered the reports as a proxy for the events and 
not studied the reports in details to extract further information, it would not be 
proper to draw any conclusions about the actors, the network relations or the 
detailed explanations/causes behind the conflicts or the violations. Certainly, it 
would be interesting to address such sociological explanations, implications and 
policies in the future.
Specific studies on the geographical implications of actor 
networks, the geographical and socio-cultural influence of their robustness under 
possible intervention will be important issues to study.

\section*{Methods}

\subsection*{Data acquisition and filtering}

GDELT Event Database~\cite{GDELT} contains the database of news articles from around 
the world in several languages, hosted through \textit{Google Cloud}. Using 
\textit{Google BigQuery}~\cite{GDELTcloud}, 
it is possible to extract data for each  event, having an unique 
time stamp, and providing the data about news about ethnic conflicts (EC) and human 
rights 
violations  (HR) happening around the world spanning over a large 
time scale. The data contains information about a pair of \textit{actors} involved, 
the location information of the event, as well as latitude, longitude data of actors 
and the 
event. 
We procured $45,942$ events for EC and $48,295$ for HR for a 15 year period,  
2001-2015.
We filtered out those data for which both actors were mentioned, along with their 
respective location information. 
The GDELT data has a huge fraction of missing entries. We have filtered out and 
excluded those data rows, which have at least one entry missing corresponding to the 
attributes we were interested in.
Hence, after cleaning, we analyzed $28,055$ events of EC and $36,470$ of HR. Details of the Cameo codes used for filtering the data (along with examples from the CAMEO codebook~\cite{GDELTcodebook}), data attributes and cleaning are provided in Supplementary 
Information.

\subsection*{Network construction}
Given a period of time $T$, we construct the network of `connected' actors in the 
following way: any two actors $A_1$ and $A_2$ mentioned together in an event $E_1$ 
reported at time $t \in [t_0:t_0+T]$ are `connected' by a link of unit weight,where 
$t_0$ is the beginning of an interval of time span $T$.
If another event $E_2$ within the same time window involves actors $A_2$ and $A_3$, 
then $A_3$ is connected to $A_2$ with a link of unit weight. Thus $A_1$ and 
$A_3$ are both connected to $A_2$ (see Fig.~\ref{fig:netviz}a). Aggregating 
all 
such events over the time window $T$, connected components emerge, with link weights 
increasing if the same pair of actors linking them appear in multiple events (actor 
pair mentions). 
These connected components form a complex network of nodes (actors) and links (actor 
pair mentions). An actor may be co-mentioned with several other actors and thus have 
a larger `degree', measured by the number of distinct co-actors it has. In 
principle, 
the network aggregated over a time period can have several disconnected components 
or  `clusters'. Fig.~\ref{fig:netviz}b shows the 
aggregated network for one year for actors mentioned in the EC dataset.

\begin{acknowledgements}
The authors thank L. Dey for useful comments. 
\end{acknowledgements}

% 
% 
% \section*{Author contributions statement}
% 
% A.C. and A.C. conceived the experiment. K.S, G. Sehgal and B.G. collected data. 
% K.S., G. Sehgal, B.G, A.C and A.C. analyzed data. All authors reviewed the 
% manuscript.
% 
% \section*{Additional information}
% 
% \textbf{Supplementary information}: accompanies the paper.
% 
% \noindent \textbf{Competing financial interests}:  The authors declare that they 
% have no competing interests.

%\bibliographystyle{unsrt}
%\bibliography{ref}

\onecolumngrid
\newpage
%\appendix

\numberwithin{table}{section}
\numberwithin{figure}{section}

\renewcommand{\thetable}{S\arabic{table}}   
\renewcommand{\thefigure}{S\arabic{figure}}

\begin{center}
\begin{large}
\textbf{Supplementary Information}
\end{large}
\end{center}

\section*{Data Description}
\label{sec:sidata}

The data source for this quantitative analysis on the global structure of ethnic 
violence is GDELT, the Global Database of Events, Language, and Tone~\cite{GDELT}. 
It is described~\cite{leetaru2013gdelt}  as ``an initiative to construct a catalog 
of human societal-scale behavior and beliefs across all countries of the world, 
connecting every person, organization, location, count, theme, news source, and 
event 
across the planet into a single massive network that captures what's happening 
around 
the world, what its context is and who's involved, and how the world is feeling 
about 
it, every single day".

GDELT contains data since 1979~\cite{stuster2013mapped}. However, the data 
from the year 2000 onwards is more comprehensive, reflecting the increase in the 
number of news media and the frequency of event recording. The entire GDELT dataset 
is available as a public dataset in Google Big Query~\cite{GDELTcloud}. GDELT event 
records are stored in an expanded version of the dyadic CAMEO format, capturing two 
actors and the action performed by \textit{Actor1} upon \textit{Actor2}. A wide 
array 
of variables break out the raw CAMEO actor codes into their respective fields to 
make 
it easier to interact with the data, the \textit{Action} codes are broken down into 
a hierarchical structure, a score describing the intensity of conflict or 
cooperation is provided, an average \textit{tone} score is provided for all coverage 
of the event, several indicators of ``importance" based on media attention are 
provided, and an unique array of geo-referencing fields offer estimated 
landmark-centroid-level geographic positioning of both actors and the location of 
the 
action.

However, for the purpose of this quantitative analysis we have extracted the data 
for 
the cameo codes 
203 (engage in ethnic cleansing) for Ethnic Conflicts (EC) and 
092 (investigate human rights abuses), 1122 (accuse of human rights abuses) for 
Human Rights Violations (HR) 
from the years 2001 to 2015. 

As mentioned in the CAMEO codebook~\cite{GDELTcodebook}, for the corresponding entries, we reproduce below the cited examples:
\begin{itemize}
\item \textbf{\texttt{Cameo code 203} for EC:} {\color{red} Serb forces} {\color{green}were engaged in ethnic cleansing} in Kosovo {\color{green}against} the majority {\color{blue} Albanian population} of the province, according to the US government.

\item \textbf{\texttt{Cameo code 092} for HR:} {\color{red} Israel}'s high court {\color{green} opened} a landmark {\color{green} hearing} Wednesday into the legality of secret {\color{green} interrogation techniques} used against {\color{blue} Palestinian} detainees.

\item \textbf{\texttt{Cameo code 1122} for HR:} Human rights watchdog {\color{red} Amnesty International} {\color{green} accused} the {\color{blue} United States} of {\color{green} violating human rights}, ignoring international law and sending a ``permissive signal to abusive governments''.
\end{itemize}

Each data entry with a \textit{GLobalEventID} had the following 
attributes~\cite{GDELTcodebook}:
\begin{itemize}
 \item 
\texttt{Actor1Code, Actor2Code}: The complete raw CAMEO code for \textit{Actor1} and 
\textit{Actor2} (includes geographic, class, ethnic, religious, and type classes). 
It may be blank if the system was unable to identify any actor.

\item
\texttt{Actor1Name, Actor2Name}: The actual name of the \textit{Actor1} and 
\textit{Actor2}. In the case of a political leader or organization, this will be the 
leader's formal name (e.g., GEORGE W BUSH, UNITED NATIONS), for a geographic match 
it will be either the country or capital/major city name (e.g., UNITED STATES / 
PARIS), and for ethnic, religious, and type matches it will reflect the root match 
class (e.g., KURD, CATHOLIC, POLICE OFFICER, etc). It may be blank if the system was 
unable to identify an actor.

\item
\texttt{Actor1Geo\_ADM1Code, Actor2Geo\_ADM2Code}: This is the 2-character FIPS10-4 
country code followed by the 2-character FIPS10-4 administrative division 1 (ADM1) 
code for the administrative division housing the landmark.

\item
\texttt{ActionGeo\_FullName}: Location of Event.  This is the full human-readable 
name of the matched location.

\item
\texttt{ActionGeo\_CountryCode}: Location of Event. This is the 2-character FIPS10-4 
country code for the location.

\item
\texttt{ActionGeo\_ADM1Code}: Location of Event.  This is the 2-character FIPS10-4 
country code followed by the 2-character FIPS10-4 administrative division 1 (ADM1) 
code for the administrative division housing the landmark.

\item
\texttt{ActionGeo\_Lat, ActionGeo\_Long}: This is the centroid latitude and 
longitude 
of the landmark for mapping.

\item 
\texttt{SQLDATE}: Date the event took place in \textit{YYYYMMDD} format.

\end{itemize}

There may be actors with similar \texttt{ActorCode} at different locations. So, for 
uniquely identifying each actor we have concatenated \texttt{Actor1Code} with 
\texttt{Actor1Geo\_ADM1Code} and \texttt{Actor2Code} with 
\texttt{Actor2Geo\_ADM2Code}. 
The data acquired from queries had to further cleaned for missing entries. 
Each event entry mentions a pair of unique actors, but we also found rare instances 
($<0.5\%$) where only one actor has been identified. 
Any row 
of data with missing actor names, actor codes or location data were removed to 
created the working data set.
The number of events were $45,942$ for EC and $48,295$ for HR 
%and $12106$ for LC,
and this was reduced to $28,055$ for EC and $36,470$ for HR 
%and $9123$ for LC 
after filtering. On actual inspection of the data, we found that a very small 
fraction of entries do not actually report an actual case of ethnic conflict or 
human rights violations, yet contain the relevant keywords that dicusss the issues 
in a positive tone (e.g. absence of ethnic violence or human right violations etc.).

\section*{The structure of the network}
% 
% %%%%%%%%%%%%%%%%%%%%%%%%%%%

%%%%%%%%%%%%%%%%%%%%%%%%%%%%%%%%%%%
\begin{figure}[H]
\centering
\includegraphics[width=16.0cm]{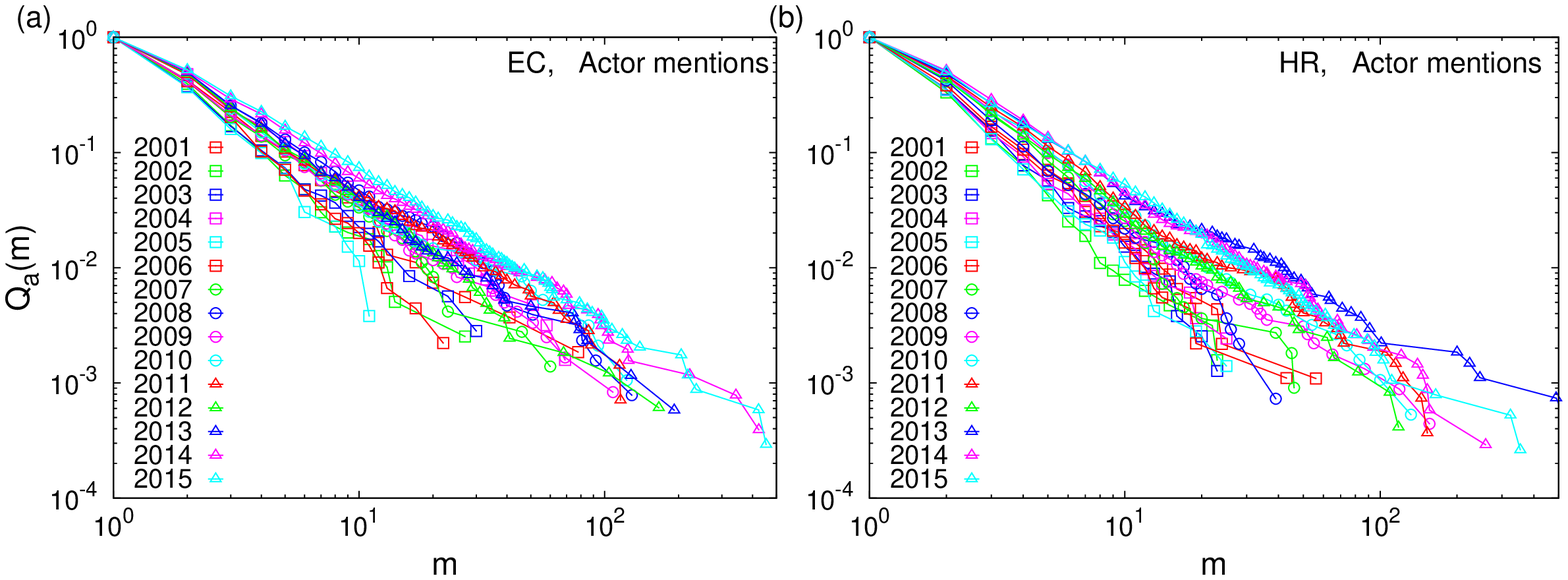}
\includegraphics[width=16.0cm]{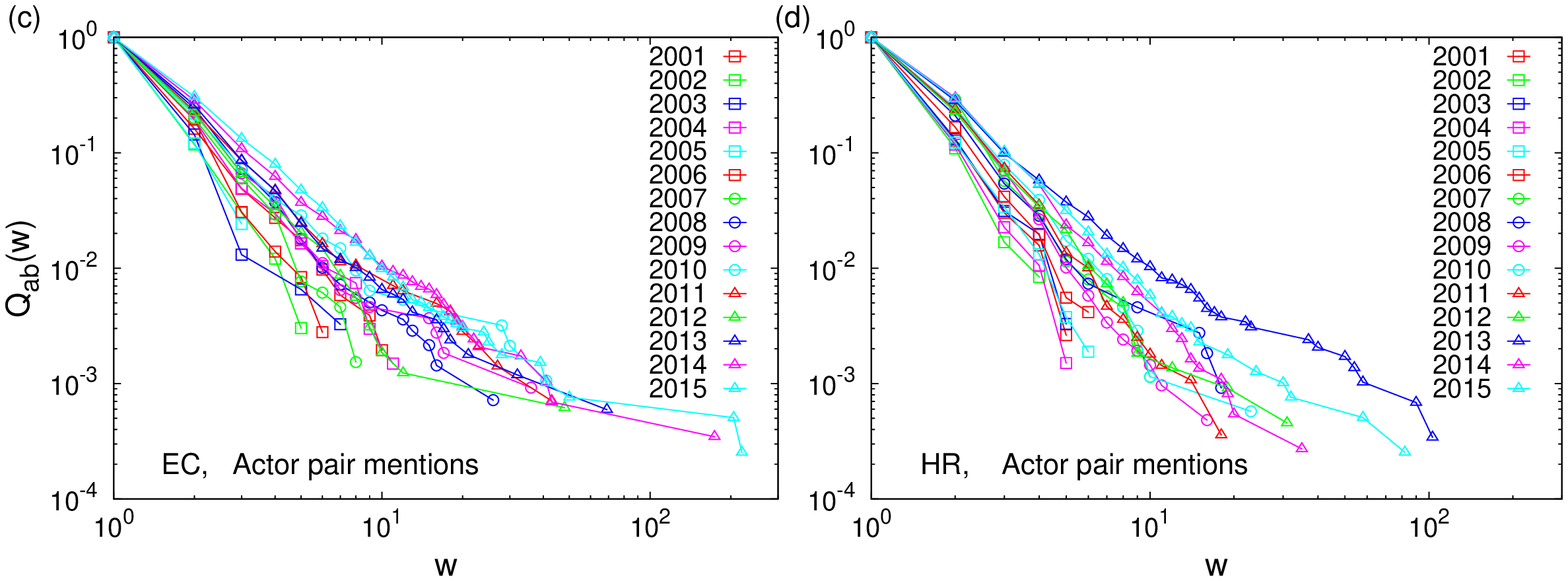}
\includegraphics[width=16.0cm]{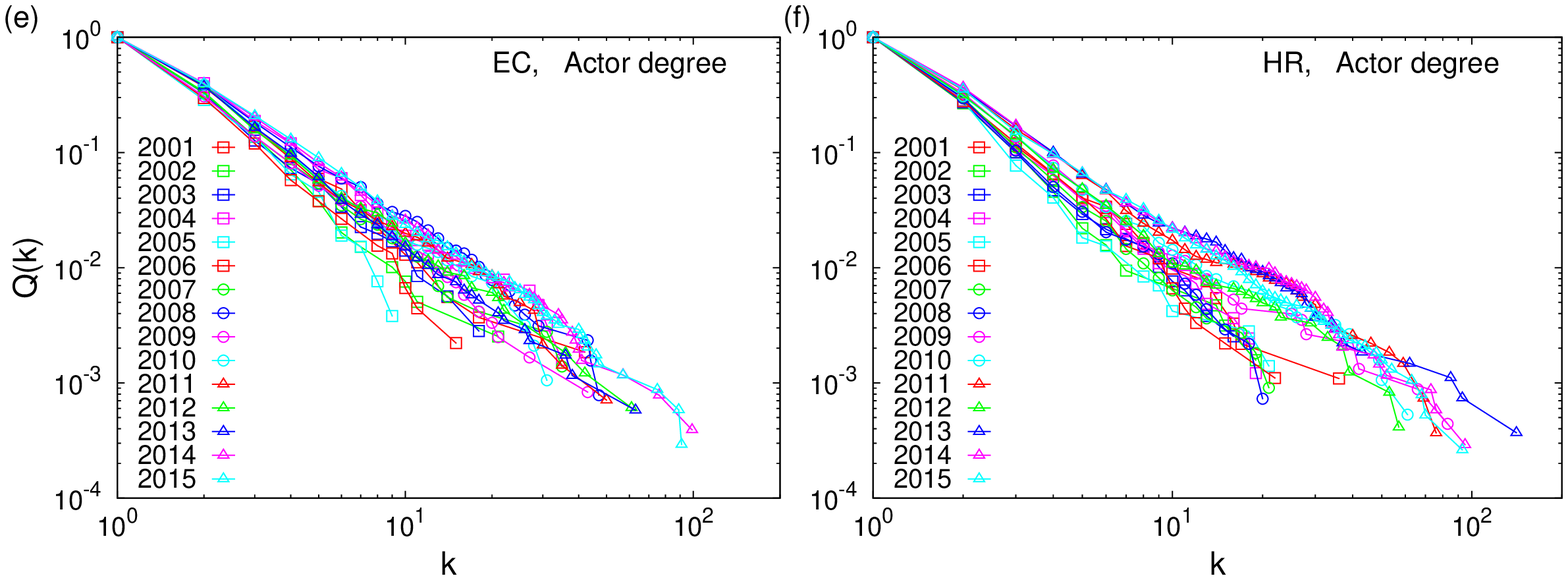}
\caption{Plots of the cumulative probability (CCDF) $Q_a(m)$ an actor is 
mentioned $m$ times or more, for (a) EC and (b) HR;
$Q_{ab}(w)$ that an actor pair is mentioned $w$ times or more for (c) EC and (d) HR, 
and $Q(k)$ that an actor is co-mentioned with $k$ actors or more for (e) EC and (f) 
HR,  for each year in the period 2001-2015.
%Guiding lines for $x^{-1.5}$ and $x^{-2.0}$ are provided and 
The actual fits are 
given in Table~\ref{tab:fits}.
}
\label{fig:yr_count}
\end{figure}
%%%%%%%%%%%%%%%%%%%%%%%%%%%

%%%%%%%%%%%%%%%%%%%%%%%%%%%%%%%%%%%
\begin{figure}[H]
\centering
\includegraphics[width=16.0cm]{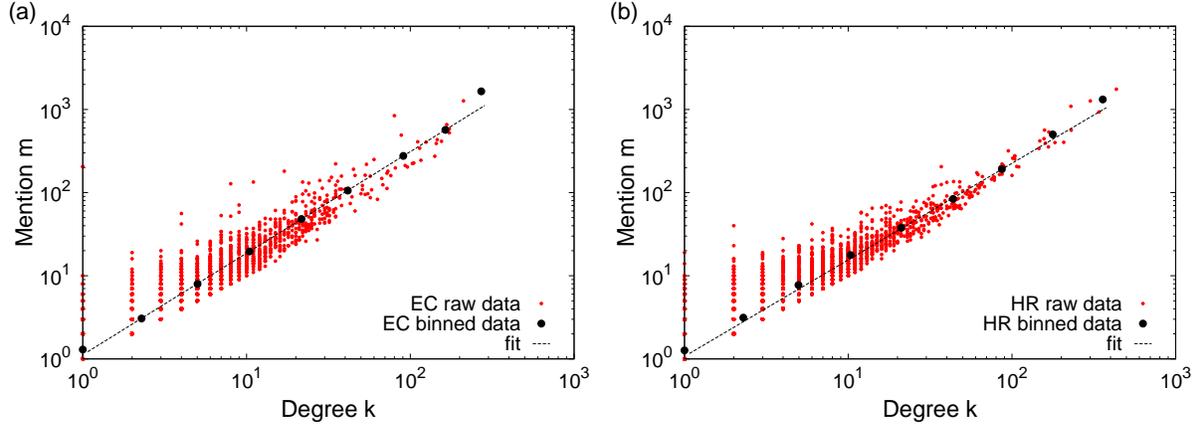}
\caption{Scatter plots showing the degree ($k$)  versus mention ($m$) (red) for 
each actor, in the networks for (a) EC and (b) HR. The raw data (red) are then 
log-binned and plotted as black filled circles. The dashed lines are the best fits 
obtained using ordinary least squares for the log-binned data (black filled 
circles): $m \propto k^\gamma$ with $\gamma$ measured to be $1.22\pm 0.01$ for EC 
and 
$1.16\pm0.02$ for HR. 
The slope of each best fit line implies strong correlation between the degree ($k$)  
and mention ($m$) for actors, as also reflected by the ratios of the power-law 
exponents of the probability distributions of the respective variables in 
Fig.~\ref{fig:allyr_cumcount}: $\nu_3/\nu_1 \approx 1.23$ for EC and $\approx 1.16$ 
for HR, compares well to corresponding values of $\gamma$.
}
\label{fig:scatterplots_logbinned}
\end{figure}
%%%%%%%%%%%%%%%%%%%%%%%%%%%

%%%%%%%%%%%%%%%%%%%%%%%%%%%%%%%%
\begin{table}[H]
\caption{Computed exponents power law distribution fits 
for (i)  actor mentions, $\nu_1$ (ii) actor pair mentions, $\nu_2$  and (iii) 
degree, $\nu_3$ 
from the data for 
Ethnic Conflicts (EC) and  Human Rights Violations (HR) for
different years (2001-2015) as well as for the 15 year aggregate.
}
\label{tab:fits}
%\begin{adjustbox}{width=\linewidth, height=1.5 in}
%\scriptsize
\centering
\begin{tabular}{|c|c|c|c|c|c|c|}
\hline
\multirow{2}{*}{Year} & \multicolumn{2}{c|}{Actor pair mention} & 
\multicolumn{2}{c|}{Actor mention} & \multicolumn{2}{c|}{Actor degree} \\ 
\cline{2-7} 
                      & EC          & HR                & EC         & HR     
        & EC        & HR             \\ \hline

2001                       & 2.61 $\pm$  0.07           & 3.47 $\pm$  0.35      
              & 1.56 $\pm$  0.05           & 1.79 $\pm$ 
 0.04                      & 2.13 $\pm$  0.05           & 
2.01 $\pm$  0.06                    \\ \hline
2002                       & 3.47 $\pm$  0.23           & 3.55 $\pm$  0.22      
               & 1.90 $\pm$  0.06           & 2.06 $\pm$ 
 0.07                      & 2.44 $\pm$  0.07           & 
2.16 $\pm$  0.09                   \\ \hline
2003                       & 3.04 $\pm$  0.37           & 3.32 $\pm$  0.35      
               & 1.71 $\pm$  0.03           & 2.03 $\pm$ 
 0.07                     & 2.01 $\pm$  0.05           & 
2.15 $\pm$  0.04                    \\ \hline
2004                       & 2.53 $\pm$  0.11           & 3.81 $\pm$  0.39     
                & 1.42 $\pm$  0.02           & 1.90 
$\pm$  0.04                    & 1.66 $\pm$  0.04       
    & 2.06 $\pm$  0.03                \\ \hline
2005                       & 3.35 $\pm$  0.24           & 3.48 $\pm$  0.18     
               & 2.15 $\pm$  0.11           & 2.08 
$\pm$  0.05                    & 2.29 $\pm$  0.15       
    & 2.16 $\pm$  0.10                   \\ \hline
2006                       & 3.25 $\pm$  0.20           & 3.14 $\pm$  0.17     
               & 2.15 $\pm$  0.07           & 1.99 
$\pm$   0.08                    & 2.24 $\pm$  0.09       
    & 2.30 $\pm$  0.07                     \\ \hline
2007                       & 3.04 $\pm$  0.18           & 3.04 $\pm$  0.14     
               & 1.64 $\pm$  0.04          & 1.86 
$\pm$  0.07                    & 1.85 $\pm$  0.05       
    & 2.25 $\pm$  0.05                 \\ \hline
2008                       & 2.27 $\pm$  0.07          & 2.28 $\pm$  0.13     
               & 1.43 $\pm$  0.03           & 1.74 
$\pm$  0.04                      & 1.69 $\pm$  0.03       
    & 2.11 $\pm$   0.05                 \\ \hline
2009                       & 2.07 $\pm$   0.17          & 2.97 $\pm$  0.09     
               & 1.50 $\pm$  0.02           & 1.54 
$\pm$  0.03                     & 1.93 $\pm$  0.04       
    & 1.69 $\pm$  0.05                   \\ \hline
2010                       & 1.47 $\pm$  0.10           & 2.61 $\pm$  0.15     
                & 1.30 $\pm$  0.03           & 1.49 
$\pm$  0.02                      & 1.71 $\pm$  0.03       
    & 1.62 $\pm$  0.04                   \\ \hline
2011                       & 1.67 $\pm$  0.07           & 2.81 $\pm$  0.07     
                 & 1.32 $\pm$  0.01           & 1.38 
$\pm$  0.03                   & 1.62 $\pm$  0.04       
    & 1.56 $\pm$  0.03                   \\ \hline
2012                       & 2.86 $\pm$  0.13           & 2.60 $\pm$  0.13     
                & 1.54 $\pm$  0.02           & 1.52 
$\pm$  0.02                     & 1.80 $\pm$   0.03      
    & 1.72 $\pm$  0.04                 \\ \hline
2013                       & 1.92 $\pm$  0.07           & 1.41 $\pm$  0.05     
             & 1.40 $\pm$  0.02           & 1.21 
$\pm$  0.02                   & 1.86 $\pm$  0.02       
    & 1.57 $\pm$  0.02                \\ \hline
2014                       & 1.83 $\pm$  0.04           & 2.48 $\pm$  0.05     
                & 1.29 $\pm$  0.01           & 1.37 
$\pm$  0.01                     & 1.69 $\pm$  0.02       
    & 1.60 $\pm$  0.03                    \\ \hline
2015                       & 1.82 $\pm$  0.06           & 2.06 $\pm$  0.07     
               & 1.31 $\pm$  0.01           & 1.45 
$\pm$  0.01                    & 1.68 $\pm$  0.02       
    & 1.64 $\pm$  0.01                   \\ \hline \hline 
2001-2015                    & 1.58 $\pm$  0.01           & 1.74 $\pm$  0.03      
                 & 1.23 $\pm$  0.01           & 1.28 
$\pm$   0.01                  & 1.52 $\pm$  0.01       
    & 1.48 $\pm$  0.01                  \\ \hline
\end{tabular}
%\end{adjustbox}
\end{table}

%%%%%%%%%%%%%%%%%%%%%%%%%%%%%%%%%%%%%%%%%%%%%%%%%%%%%%%%%%%%%%%%%%%%
%\clearpage 
\section*{Clusters}

%%%%%%%%%%%%%%%%%%%%%%%%%%%%%%%%
\begin{table}[H]
\centering
\caption{Table showing the number of actors and the actors in the largest connected 
cluster for the different data sets: 
Ethnic Conflicts (EC) and  Human Rights Violations (HR) 
for different years (2001-2015) as well as for the 15 year aggregate.
}
\label{tab:clusfits}
\begin{tabular}{|c|c|c|c|c|}
\hline
\multirow{2}{*}{Year} & \multicolumn{2}{c|}{EC}                                      
 
                & \multicolumn{2}{c|}{HR}

    \\ \cline{2-5} 
                      & Total nodes & \begin{tabular}[c]{@{}c@{}}Largest \\ cluster 
size\end{tabular} & Total nodes & \begin{tabular}[c]{@{}c@{}}Largest \\ cluster 
size\end{tabular}  \\ \hline

2001 &	540 &	193 &	917 &	333 \\ \hline
2002 &	396 &	104 &	636 &	72   \\ \hline
2003 &	355 &	117 &	788 &	184  \\ \hline
2004 &	633 &	329 &	823 &	285 \\ \hline
2005 &	263 &	48 &	714 &	110   \\ \hline
2006 &	451 &	56 &	907 &	183   \\ \hline
2007 &	721 &	288 &	1,102 &	309 \\ \hline
2008 &	1,279 &	696 &	1,375 &	451  \\ \hline
2009 &	1,204 &	544 &	2,267 &	1,011  \\ \hline
2010 &	951 &	474 &	1,887 &	954 \\ \hline
2011 &	1,412 &	746 &	2,719 &	1,591  \\ \hline
2012 &	1,644 &	899 &	2,423 &	1,142  \\ \hline
2013 &	1,728 &	892 &	2,707 &	1,576  \\ \hline
2014 &	2,560 &	1,620 &	3,447 &	2,152  \\ \hline
2015 &	3,432 &	2,229 &	3,816 &	2,322  \\ \hline \hline  
2001-2015 &	10,394 &	7,875 &	15,899 &	12,106  \\ \hline

\end{tabular}
\end{table}

\begin{figure}[H]
\centering \includegraphics[width=10.0cm]{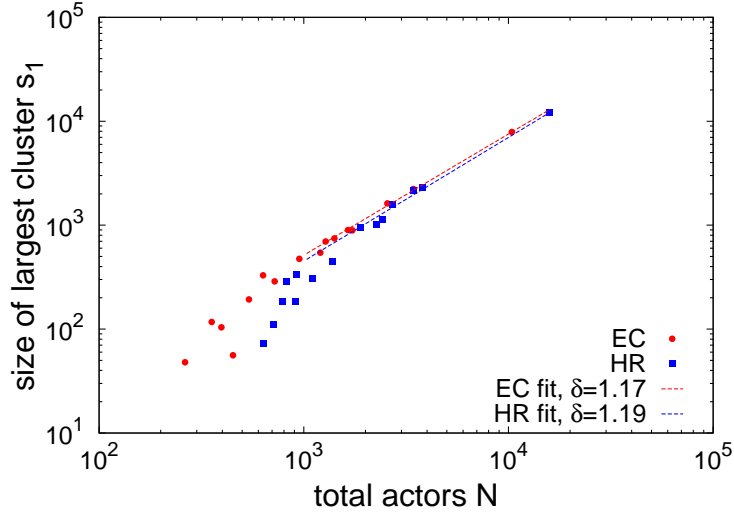}
\caption{Variation of the size of the largest cluster $s_1$ with network size $N$, 
for different years (2001-2015) and well as 15 year aggregate for EC and HR.
The power law fits to $s_1 \sim N^\delta$ are $\delta=1.17 \pm 0.01$ for EC and 
$1.19 \pm 
0.02$ for HR.}
\label{fig:clus_net}
\end{figure}

The size of the largest cluster is computed as the maximum number of nodes $s_1$ in 
the giant component or largest subgraph and shown in 
Supplementary Table.~\ref{tab:clusfits} and the variation of $s_1$ with the size of 
the entire network is shown in Fig.~\ref{fig:clus_net}, and the asymptotic fit is 
found to be $s_1 \sim N^\delta$. The degree distribution of 
the largest cluster / giant component has a power law tail. The power law exponents 
for the asymptotic fits are given in Supplementary 
Table.~\ref{Big_cluster_degree:fits}.

%%%%%%%%%%%%%%%%%%%%%%%%%%%
%%%%%%%%%%%%%%%%%%%%%%%%%%%

%%%%%%%%%%%%%%%%%%%%%%%%%%%
\begin{figure}[H]
\includegraphics[width=17.0cm]{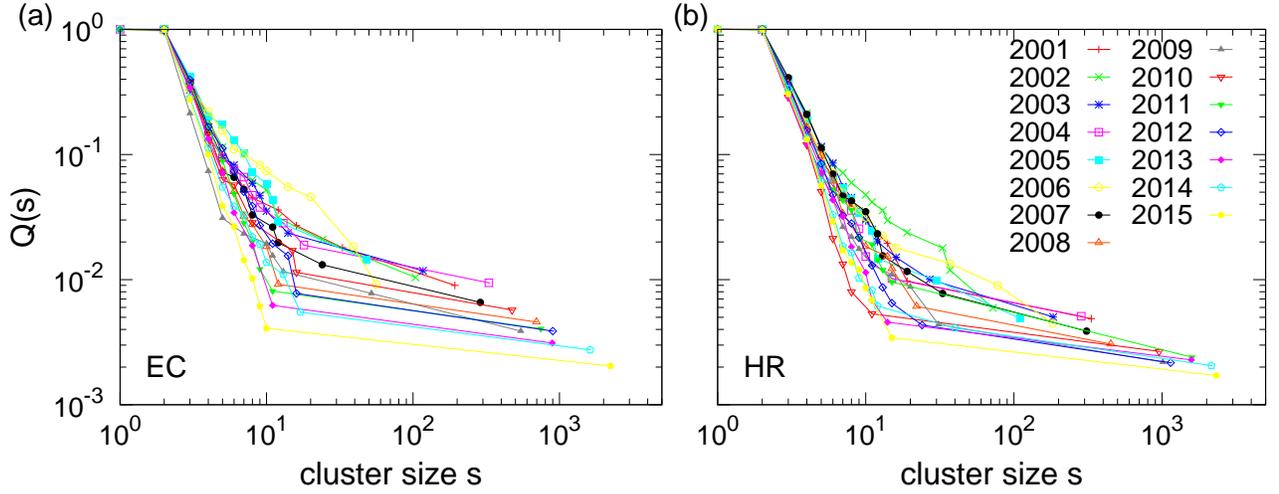}
\caption{Plots of the cumulative probability (CCDF) $Q(s)$ that there is a cluster 
of 
size 
larger than $s$, for different years (2001-2015) for (a) EC and (b) HR.}
\label{fig:clus_size3}
\end{figure}
%%%%%%%%%%%%%%%%%%%%%%%%%%%

%===================================================================
\begin{table}[H]
\centering
\caption{Computed exponents $\nu_3$ of power law distribution to asymptotic fits 
to the degree distribution $Q(k) \sim k^{-\nu_3}$
for the largest connected cluster (giant component) for the two categories 
of data -- Ethnic 
Conflicts (EC) and Human Rights Violations (HR) for individual years and the 
15 
year aggregate (2001-2015).}

%\begin{adjustbox}{width=\linewidth}
\label{Big_cluster_degree:fits}
\begin{tabular}{|l|l|l|}
\hline
\multicolumn{1}{|c|}{Year} & \multicolumn{1}{c|}{EC} & \multicolumn{1}{c|}{HR}  \\ 
\hline
2001                       & 1.99 $\pm$  0.06           & 1.71 $\pm$  0.07    
                 \\ \hline
2002                       & 1.84 $\pm$  0.08           & 1.65 $\pm$  0.05    
                 \\ \hline
2003                       & 1.84 $\pm$  0.08           & 1.56 $\pm$  0.05    
                  \\ \hline
2004                       & 1.58 $\pm$   0.05          & 1.83 $\pm$  0.06    
                 \\ \hline
2005                       & 1.75 $\pm$  0.29           & 1.48 $\pm$  0.09    
                 \\ \hline
2006                       & 1.50 $\pm$  0.05           & 1.97 $\pm$  0.08    
                 \\ \hline
2007                       & 1.71 $\pm$  0.06           & 1.84 $\pm$  0.04    
                 \\ \hline
2008                       & 1.64 $\pm$  0.04           & 1.81 $\pm$  0.08    
                 \\ \hline
2009                       & 1.88 $\pm$  0.05           & 1.56 $\pm$  0.04    
                \\ \hline
2010                       & 1.62 $\pm$  0.04           & 1.49 $\pm$  0.03   
                  \\ \hline
2011                       & 1.50 $\pm$   0.03          & 1.41 $\pm$  0.04   
                  \\ \hline
2012                       & 1.75 $\pm$   0.03          & 1.60 $\pm$  0.04    
                  \\ \hline
2013                       & 1.83 $\pm$  0.02           & 1.53 $\pm$  0.02    
                \\ \hline
2014                       & 1.67 $\pm$  0.03           & 1.57 $\pm$  0.04    
                  \\ \hline
2015                       & 1.69 $\pm$  0.02           & 1.60 $\pm$  0.01    
                  \\ \hline \hline 
2001-2015                    & 1.51 $\pm$  0.01           & 1.48 $\pm$  0.01    
                \\ \hline

\end{tabular}
%\end{adjustbox}
\end{table}

%============================================================
\subsection*{Biggest cluster: clustering coefficient }
\begin{figure}[H]
\centering 
\includegraphics[width=8.5cm]{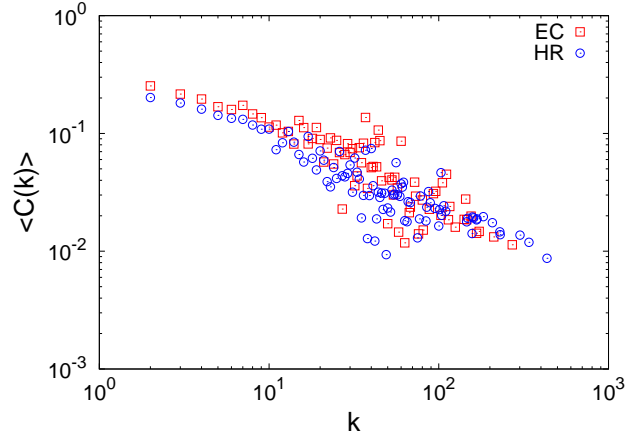}
\caption{Plot of average clustering coefficient $\langle C(k) \rangle$ with degree 
$k$ of the biggest cluster for the two categories of data -- EC and HR for 
the 15 years aggregate (2001-2015). The data for all sets exhibit a slow decay with 
degree.}
\label{fig:Big_Clus_CC}
\end{figure}

To find the cohesion of ethnic conflict and human rights violation, we computed the 
clustering coefficient of both (5 years 
aggregated) networks, which shows the extents to which the nodes of a network are 
closely connected with one another.
The clustering coefficient for a node $i$ in an undirected graph is computed as:
$C_i= \frac{2 N_i}{k_i(k_i-1)}$,
where $k_i$ is the degree of the node $i$, and $N_i$ is the number of 
links between the neighbors of $i$.
We plot the average clustering coefficient of a node with degree $k$ in 
Fig.~\ref{fig:Big_Clus_CC}, 
and find that higher degree nodes are less  clustered compared to low degree nodes.

%%%%%%%%%%%%%%%%%%%

%============================================================

%\clearpage 
\subsection*{Dynamics of network growth}

%%%%%%%%%%%%%%%%%%%%%%%%%%%%%%%
\begin{figure*}[h]
\includegraphics[width=8.5cm]{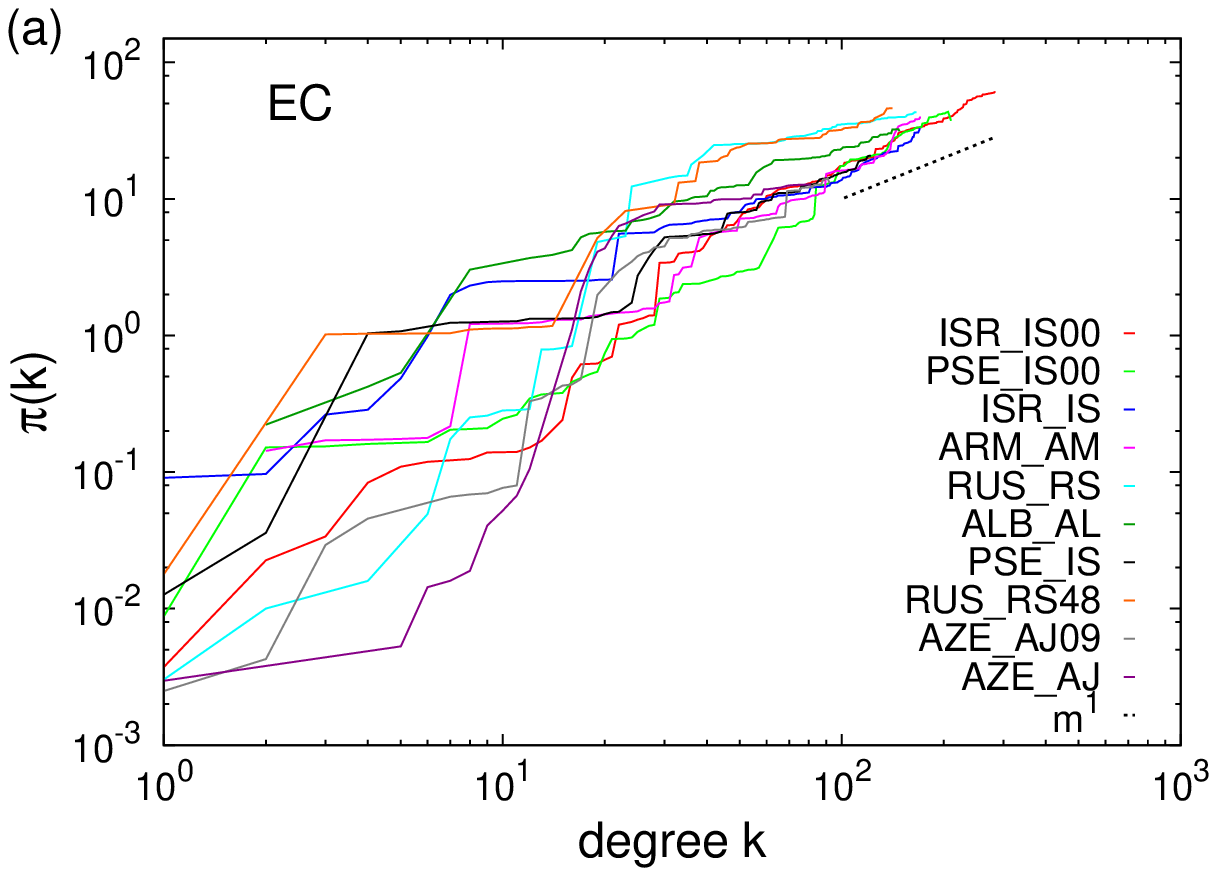}
\includegraphics[width=8.5cm]{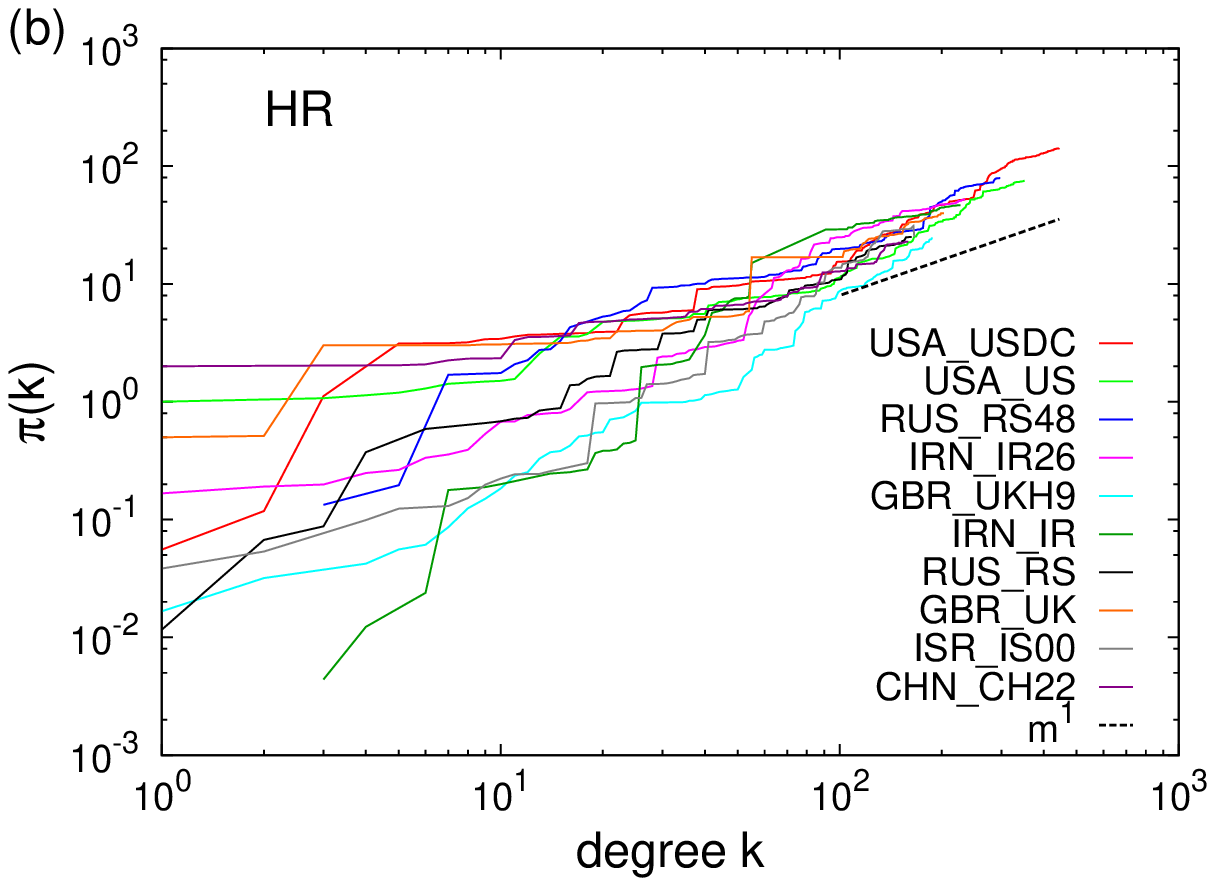}
\caption{Plots showing the cumulative growth rates $\pi(k)$ for degree $k$ for (A) EC 
and (B) HR 
datasets. The 
curves fit to  $\pi(k) \sim  k^\alpha$ with $\alpha > 1$. The precise fitting 
exponents are 
given in  Table.~\ref{tab:fitsgrow}.}
\label{fig:pik_deg}
\end{figure*}
%%%%%%%%%%%%%%%%%%%%%%%%%%%%%%%%

\begin{table}[H]
 \centering
\caption{Computed exponents of asymptotic power law  fits 
for the cumulative growth rates $\pi(x) \sim x^\alpha$
for the top ten highest counts for (i)  actor mentions, (ii)  degree, 
for the 2 categories of data -- 
Ethnic Conflicts (EC) and Human Rights Violations (HR) for the 15 year aggregate.
}
\label{tab:fitsgrow}

\begin{tabular}{|l|l|l|l|l|l|}
\hline
\multicolumn{3}{|c|}{EC}                                                             
 
   & \multicolumn{3}{c|}{HR}                                                         
 
       \\ \hline
\multicolumn{1}{|c|}{Actor} & \multicolumn{1}{c|}{Degree} & 
\multicolumn{1}{c|}{Mention} & \multicolumn{1}{c|}{Actor} & 
\multicolumn{1}{c|}{Degree} & \multicolumn{1}{c|}{Mention} \\ \hline
ALB\_AL                     & 0.86 $\pm$  0.01               & 1.04 $\pm$  
0.01                & CHN\_CH22                  & 0.71 $\pm$  0.02             
  & 1.11 $\pm$  0.01                \\ \hline
ARM\_AM                     & 1.55$\pm$  0.02               & 1.51$\pm$  0.01 
               & GBR\_UK                    & 1.11$\pm$  0.02               & 
1.33$\pm$  0.01                \\ \hline
AZE\_AJ                     & 0.50$\pm$  0.02               & 0.96$\pm$  0.01 
               & GBR\_UKH9                  & 1.68$\pm$  0.02               & 
1.79$\pm$  0.01                \\ \hline
AZE\_AJ09                   & 0.96$\pm$  0.04               & 1.28$\pm$  0.01 
               & IRN\_IR                    & 0.57$\pm$  0.01               & 
1.10$\pm$  0.01                \\ \hline
ISR\_IS                     & 0.89$\pm$  0.02               & 1.23$\pm$  0.01 
               & IRN\_IR26                  & 1.58$\pm$  0.03               & 
1.47$\pm$  0.01                \\ \hline
ISR\_IS00                   & 1.26$\pm$  0.01               & 1.38$\pm$  0.01 
               & ISR\_IS00                  & 1.86$\pm$  0.02               & 
1.71$\pm$  0.01                \\ \hline
PSE\_IS                     & 1.05$\pm$  0.02               & 1.31$\pm$  0.01 
               & RUS\_RS                    & 1.27$\pm$  0.02               & 
1.41$\pm$  0.01                \\ \hline
PSE\_IS00                   & 1.82$\pm$  0.02               & 1.40$\pm$  0.01 
               & RUS\_RS48                  & 1.04$\pm$  0.02               & 
1.22$\pm$  0.01                \\ \hline
RUS\_RS                     & 0.47$\pm$  0.01               & 0.92$\pm$  0.01 
               & USA\_US                    & 1.33$\pm$  0.02               & 
1.40$\pm$  0.01                \\ \hline
RUS\_RS48                   & 0.53$\pm$  0.02               & 0.79$\pm$  0.01 
               & USA\_USDC                  & 1.29$\pm$  0.01               & 
1.35$\pm$  0.01                \\ \hline
\end{tabular}
\end{table}

\section*{Measuring causality} 
Let us consider the two random variables depicting counts of EC (Ethnic Conflicts 
mentions) and HR (Human Rights Violation mentions). To say that EC causes HR, 
Granger causality~\cite{granger1969investigating} computes a regression of variable 
HR on the past values of itself 
and the past values of EC and then tests the significance of coefficient estimates 
associated with EC.

We consider a bivariate linear autoregressive model on EC and HR, and assume the 
L.H.S. to be dependent on the history of EC and HR,
\begin{equation}
HR_t  =  a_0 + a_1 EC_{t-1}+ \ldots + a_h EC_{t-h} + b_1 HR_{t-1}+ \ldots +b_h 
HR_{t-h} + E_t
\end{equation}
where $h$ is the maximum number of lagged observations (for both EC and HR). The 
coefficients $a_i$, $b_i$ are the contributions of each lagged observation to the 
predicted value of $EC_{t-i}$ and $HR_{t-i}$ respectively while $E_t$ is the 
prediction error.

If $b_1=b_2=\ldots=b_h =0$, we call it a null hypothesis $HR_0$ which implies that 
EC does not cause HR. In other words the coefficients of EC are not significant 
enough to cause HR. But if the null hypothesis gets rejected we say that the 
coefficients of EC are significant enough to cause HR.

For testing this significance of the coefficients, we compute the $p$-value. If the 
$p$-value is less that $0.05$ one can reject the null hypothesis, and hence 
conclude that EC causes HR (HR $\sim$ EC).

Applying the above process on the EC and HR \textit{year wise} mentions, we found 
that:
\begin{itemize}
 \item 
On testing $EC \sim HR$ for $h=4$, we find $p \simeq 0.721$, and  the null 
hypothesis 
cannot be rejected. Hence, one cannot conclude that HR causes EC.

\item On testing $HR \sim EC$ for $h=4$, we find $p \simeq 0.029$, and the null 
hypothesis can be rejected. Hence, we can say that EC causes HR.
\end{itemize}

Applying the above process on the EC and HR \textit{month wise} mentions, we found 
that:
\begin{itemize}
 \item 
On testing $EC \sim HR$ for $h=5$, we find $p \simeq 0.193$ and thus the null 
hypothesis 
cannot be rejected. Hence, we cannot say that HR causes EC.

\item On testing $HR \sim EC$ for $h=5$, we find $p\simeq 0.036$, and thus the 
null 
hypothesis can be rejected. Hence, we can say that EC 	causes HR.
\end{itemize}

Hence, we can definitely conclude from the above quantitative analysis that Ethnic 
conflicts cause Human Rights violations.

\end{document}